\documentclass[english]{article}
\usepackage[T1]{fontenc}
\usepackage[latin9]{inputenc}
\usepackage{geometry}
\usepackage{color}
\usepackage{mathtools}
\usepackage{amsmath}
\usepackage{amssymb}
\usepackage{esint}
\usepackage{babel}
\begin{document}

\title{\textcolor{blue}{\huge{}Semiclassical Newtonian Field Theories Based
On Stochastic Mechanics II}}

\author{Maaneli Derakhshani\thanks{Email: maanelid@yahoo.com and m.derakhshani@uu.nl}}

\maketitle
\begin{center}
\emph{\large{}Institute for History and Foundations of Science \&
Department of Mathematics, Utrecht University, Utrecht, The Netherlands}
\par\end{center}{\large \par}

~
\begin{abstract}
Continuing the development of the ZSM-Newton/Coulomb approach to semiclassical
Newtonian gravity/electrodynamics \cite{Derakhshani2017}, we formulate
a ZSM-Newton/Coulomb version of the large N approximation scheme proposed
by Oriols et al. \cite{Oriols2016}. We show that this new large N
scheme makes it possible to self-consistently describe the center-of-mass
evolution of a large number of gravitationally/electrostatically interacting,
identical, \emph{zbw} particles, without assuming that the particles
are weakly coupled, and without entailing the problematic macroscopic
semiclassical gravitational/electrostatic cat states characteristic
of the mean-field Schr{\"o}dinger-Newton/Coulomb equations. We also show
how to recover N-particle classical Newtonian gravity/electrodynamics
for many gravitationally/electrostatically interacting macroscopic
particles (composed of many interacting \emph{zbw} particles), as
well as classical Vlasov-Poisson mean-field theory for macroscopic
particles weakly interacting gravitationally/electrostatically. Finally,
we outline an explicit model of environmental decoherence that can
be incorporated into Oriols et al.'s scheme as applied to ZSM-Newton/Coulomb.
\end{abstract}

\section{Introduction}

This paper is a direct continuation of Part I \cite{Derakhshani2017}.
There, we formulated fundamentally-semiclassical Newtonian gravity/electrodynamics
based on stochastic mechanics in the ZSM formulation (ZSM-Newton/Coulomb).
Our key results were: (i) ZSM-Newton/Coulomb has a consistent statistical
interpretation; (ii) ZSM-Newton/Coulomb recovers the standard quantum
description of non-relativistic matter-gravity/charge-field coupling
as a special case valid for all practical purposes, even though the
gravitational/electrostatic interaction between \emph{zbw} particles
is fundamentally classical; and (iii) ZSM-Newton/Coulomb recovers
the `single-body' Schr{\"o}dinger-Newton/Coulomb (SN/SC) and stochastic
SN/SC equations as mean-field approximations for systems of gravitationally/electrostatically
interacting, identical, \emph{zbw} particles, in the weak-coupling
large N limit. 

We also discussed some limitations of the mean-field SN/SC and stochastic
SN/SC equations: (i) they are based on the assumption that interactions
between \emph{zbw} particles are sufficiently weak that the independent
particle approximation is plausible; and (ii) the single-body SN/SC
and stochastic SN/SC equations admit solutions corresponding to macroscopic
semiclassical gravitational/electrostatic cat states, and these cat
states predict unphysical gravitational/electrostatic forces on external
probe masses. (We also pointed out that the latter difficulty afflicts
any formulation of fundamentally-semiclassical Newtonian gravity/electrodynamics
based on the many-body SN/SC and stochastic SN/SC equations, as these
equations also allow for such cat states.)

The primary objective of the present paper is to develop a new large
N scheme for ZSM-Newton/Coulomb, that bypasses the limitations of
the mean-field SN/SC and stochastic SN/SC equations.

Our scheme will be based on the one developed recently by Oriols et
al. \cite{Oriols2016}, who consider the center-of-mass (CM) motion
of a system of N identical, non-relativistic, de Broglie-Bohm (dBB)
particles coupled through interaction potentials of the form $\hat{U}_{int}(\hat{x}_{j}-\hat{x}_{k})$
and to external potentials of the form $\hat{U}_{ext}(\hat{x}_{j})$.
They show that, in the limit $N\rightarrow\infty$, the CM motion
becomes effectively indistinguishable from classical Hamilton-Jacobi
mechanics for a single massive particle in an external field. 

Essentially, we will import the Oriols et al. scheme into ZSM-Newton/Coulomb.
In doing so, we will find that it is possible to: (i) self-consistently
describe the CM motion of large numbers of classical-gravitationally/electrostatically
interacting, identical, \emph{zbw} particles, without an independent
particle approximation; (ii) avoid macroscopic semiclassical gravitational/electrostatic
cat states and recover many-particle classical Newtonian gravity/electrodynamics
for the CM descriptions of gravitationally/electrostatically interacting
macroscopic particles (where the macroscopic particles are composed
of many interacting \emph{zbw} particles); and (iii) recover classical
Vlasov-Poisson mean-field theory for macroscopic particles that interact
gravitationally/electrostatically, in the weak-coupling large particle
number limit. We will also be led to suggest an explicit model of
environmental decoherence that's consistent with the Oriols et al.
scheme, and which could justify a crucial assumption of the scheme.

The outline of the paper is as follows. Section 2 implements the Oriols
et al. scheme into ZSM-Newton/Coulomb, and shows how classical Newtonian
dynamics for the center-of-mass of a many-particle system is recovered
in the large N limit. Section 3 shows how to derive the classical
nonlinear Schr{\"o}dinger equation for the large N center-of-mass motion.
Section 4 shows how to recover classical Newtonian gravity/electrodynamics
for many gravitationally/electrostatically interacting macroscopic
particles. Section 5 shows how to recover classical Vlasov-Poisson
mean-field theory. Section 6 sketches an explicit model of environmental
decoherence that's consistent with the Oriols et al. scheme applied
to ZSM-Newton/Coulomb.

\section{Large N center-of-mass approximation in ZSM-Newton/Coulomb}

\subsection{General approach}

We begin by considering ZSM for N identical \emph{zbw} particles in
(for simplicity) 1-dimensional space, with configuration $X(t)=\left\{ x_{1}(t),...,x_{N}(t)\right\} $
and the ensemble-averaged, time-symmetric, joint \emph{zbw} phase

\begin{equation}
\begin{aligned}J(X,t) & \coloneqq\int_{\mathbb{R}^{N}}d^{N}X\rho(X,t)\int_{t_{I}}^{t_{F}}\left\{ \sum_{i=1}^{N}\frac{1}{2}\left[2mc^{2}+\frac{1}{2}m\left(Dx_{i}(t)\right)^{2}+\frac{1}{2}m\left(D_{*}x_{i}(t)\right)^{2}\right]-U(X(t),t)\right\} dt+\sum_{i=1}^{N}\phi_{i},\\
 & =\int_{\mathbb{R}^{N}}d^{N}X\rho(X,t)\int_{t_{I}}^{t_{F}}\left\{ \sum_{i=1}^{N}\left[mc^{2}+\frac{1}{2}mv_{i}^{2}(X(t),t)+\frac{1}{2}mu_{i}^{2}(X(t),t)\right]-U(X(t),t)\right\} dt+\sum_{i=1}^{N}\phi_{i}
\end{aligned}
\end{equation}
where

\begin{equation}
Dx_{i}(t)=\left[\frac{\partial}{\partial t}+\sum_{i=1}^{N}\mathbf{b}_{i}(X(t),t)\frac{\partial}{\partial x_{i}}+\sum_{i=1}^{N}\frac{\hbar}{2m}\frac{\partial^{2}}{\partial x_{i}^{2}}\right]x_{i}(t),
\end{equation}

\begin{equation}
D_{*}x_{i}(t)=\left[\frac{\partial}{\partial t}+\sum_{i=1}^{N}\mathbf{b}_{i*}(X(t),t)\frac{\partial}{\partial x_{i}}-\sum_{i=1}^{N}\frac{\hbar}{2m}\frac{\partial^{2}}{\partial x_{i}^{2}}\right]x_{i}(t).
\end{equation}
The potential $U(X(t),t)$ is assumed to take the general form 
\begin{equation}
U(X(t),t)\coloneqq\sum_{j=1}^{N}U_{ext}(x_{j}(t))+\frac{1}{2}\sum_{j,k=1}^{N(j\neq k)}U_{int}(x_{j}(t)-x_{k}(t)),
\end{equation}
and we assume the usual constraints 

\begin{equation}
\mathbf{v}_{i}\coloneqq\frac{1}{2}\left[\mathbf{b}_{i}+\mathbf{b}_{i*}\right]=\frac{1}{m}\frac{\partial S}{\partial x_{i}},
\end{equation}

\begin{equation}
\mathbf{u}_{i}\coloneqq\frac{1}{2}\left[\mathbf{b}_{i}-\mathbf{b}_{i*}\right]=\frac{\hbar}{2m}\frac{1}{\rho}\frac{\partial\rho}{\partial x_{i}}.
\end{equation}
As a consequence of (5-6), the time-reversal invariant joint probability
density $\rho(X,t)$ evolves by

\begin{equation}
\frac{\partial\rho}{\partial t}=-\sum_{i=1}^{N}\frac{\partial}{\partial x_{i}}\left(\frac{\rho}{m}\frac{\partial S}{\partial x_{i}}\right),
\end{equation}
and satisfies the normalization
\begin{equation}
\int_{\mathbb{R}^{N}}\rho_{0}(X)d^{N}X=1.
\end{equation}
The stochastic differential equations of motion for $x_{i}(t)$ take
the form

\begin{equation}
dx_{i}(t)=\left[\frac{1}{m}\frac{\partial S}{\partial x_{i}}+\frac{\hbar}{2m}\frac{1}{\rho}\frac{\partial\rho}{\partial x_{i}}\right]|_{x_{j}=x_{j}(t)}dt+dW_{i}(t),
\end{equation}

\begin{equation}
dx_{i}(t)=\left[\frac{1}{m}\frac{\partial S}{\partial x_{i}}-\frac{\hbar}{2m}\frac{1}{\rho}\frac{\partial\rho}{\partial x_{i}}\right]|_{x_{j}=x_{j}(t)}dt+dW_{i*}(t),
\end{equation}
where the $dW_{i}$ ($dW_{i*}$) are 1-dimensional Wiener processes
satisfying Gaussianity, independence of $dx_{i}(s)$ for $s\leq t$,
and variance 
\begin{equation}
E_{t}\left[dW_{i}^{2}\right]=\frac{\hbar}{m}dt.
\end{equation}
Analogous conditions apply to the backward Wiener processes. 

Note that, since we are considering the case of particle motion in
a 1-dimensional space, we can disregard the quantization condition
for (5) (we will come back to it later, though, when we consider the
case of particle motion in a 3-dimensional Euclidean space).

Now, following Oriols et al. \cite{Oriols2016}, we would like to
redefine (1) in terms of the CM position $x_{cm}(t)$ and relative
positions $\mathbf{y}(t)=\left\{ y_{2}(t),...,y_{N}(t)\right\} $
such that no cross terms arise from the Laplacians in $D$ and $D_{*}$.
As shown by Oriols et al. \cite{Oriols2016}, the coordinate transformation
\begin{equation}
x_{cm}\coloneqq\frac{1}{N}\sum_{i=1}^{N}x_{i},
\end{equation}
 
\begin{equation}
y_{j}\coloneqq x_{j}-\frac{\left(\sqrt{N}x_{cm}+x_{1}\right)}{\sqrt{N}+1},
\end{equation}
makes it possible to rewrite the N-particle Schr{\"o}dinger equation,
with potential (4), in terms of $x_{cm}$ and $\mathbf{y}=\left\{ y_{2},...,y_{N}\right\} $
without cross terms arising from the Laplacian in the Schr{\"o}dinger
Hamiltonian. Thus, applying (12-13) to (1), we obtain \footnote{The proof of this goes along the same lines as Appendix A.1 of Oriols
et al. \cite{Oriols2016}.}

\begin{equation}
\begin{aligned}J(x_{cm},\mathbf{y},t) & \coloneqq\int_{\mathbb{R}^{N}}\int_{\mathbb{R}^{N-1}}dx_{cm}d^{N-1}\mathbf{y}\rho(x_{cm},\mathbf{y},t)\int_{t_{I}}^{t_{F}}\Bigl\{ Mc^{2}+\frac{m}{4}\left[\left(\tilde{D}x_{cm}(t)\right)^{2}+\left(\tilde{D}_{cm*}x_{cm}(t)\right)^{2}\right]\\
 & +\frac{m}{4}\sum_{j=2}^{N}\left[\left(\tilde{D}y_{j}(t)\right)^{2}+\left(\tilde{D}_{*}y_{j}(t)\right)^{2}\right]-U\Bigr\} dt+\phi_{cm}+\phi_{rel}=\int_{\mathbb{R}^{N}}\int_{\mathbb{R}^{N-1}}dx_{cm}d^{N-1}\mathbf{y}\rho(x_{cm},\mathbf{y},t)\\
 & \times\int_{t_{I}}^{t_{F}}\left[Mc^{2}+\frac{1}{2}M\left(v_{cm}^{2}+u_{cm}^{2}\right)+\frac{1}{2}m\sum_{j=2}^{N}\left(v_{j}^{2}+u_{j}^{2}\right)-U\right]dt+\phi_{cm}+\phi_{rel},
\end{aligned}
\end{equation}
where $M=Nm$, the CM velocities are given by

\begin{equation}
\mathbf{v}_{cm}\coloneqq\frac{1}{2}\left[\mathbf{b}_{cm}+\mathbf{b}_{cm*}\right]=\frac{1}{m}\frac{\partial S(x_{cm},\mathbf{y}(t),t)}{\partial x_{cm}}|_{x_{cm}=x_{cm}(t)},
\end{equation}

\begin{equation}
\mathbf{u}_{cm}\coloneqq\frac{1}{2}\left[\mathbf{b}_{cm}-\mathbf{b}_{cm*}\right]=\frac{\hbar}{2m}\frac{1}{\rho(x_{cm},\mathbf{y}(t),t)}\frac{\partial\rho(x_{cm},\mathbf{y}(t),t)}{\partial x_{cm}}|_{x_{cm}=x_{cm}(t)},
\end{equation}
the relative velocities are given by

\begin{equation}
\mathbf{v}_{j}\coloneqq\frac{1}{2}\left[\mathbf{b}_{j}+\mathbf{b}_{j*}\right]=\frac{1}{m}\frac{\partial S(x_{cm}(t),\mathbf{y},t)}{\partial y_{j}}|_{\mathbf{y}=\mathbf{y}(t)},
\end{equation}

\begin{equation}
\mathbf{u}_{j}\coloneqq\frac{1}{2}\left[\mathbf{b}_{j}-\mathbf{b}_{j*}\right]=\frac{\hbar}{2m}\frac{1}{\rho(x_{cm}(t),\mathbf{y},t)}\frac{\partial\rho(x_{cm}(t),\mathbf{y},t)}{\partial y_{j}}|_{\mathbf{y}=\mathbf{y}(t)},
\end{equation}
and the transformed mean forward/backward derivatives take the form

\begin{equation}
\tilde{D}x_{cm}(t)=\left[\frac{\partial}{\partial t}+\mathbf{b}_{cm}\frac{\partial}{\partial x_{cm}}+\sum_{j=2}^{N}\mathbf{b}_{j}\frac{\partial}{\partial y_{j}}+\frac{\hbar}{2M}\frac{\partial^{2}}{\partial x_{cm}^{2}}+\sum_{j=2}^{N}\frac{\hbar}{2m}\frac{\partial^{2}}{\partial y_{j}^{2}}\right]x_{cm}(t)=\mathbf{b}_{cm},
\end{equation}

\begin{equation}
\tilde{D}_{*}x_{cm}(t)=\left[\frac{\partial}{\partial t}+\mathbf{b}_{cm*}\frac{\partial}{\partial x_{cm}}+\sum_{j=2}^{N}\mathbf{b}_{j*}\frac{\partial}{\partial y_{j}}-\frac{\hbar}{2M}\frac{\partial^{2}}{\partial x_{cm}^{2}}-\sum_{j=2}^{N}\frac{\hbar}{2m}\frac{\partial^{2}}{\partial y_{j}^{2}}\right]x_{cm}(t)=\mathbf{b}_{cm*},
\end{equation}
and

\begin{equation}
\tilde{D}y_{j}(t)=\mathbf{b}_{j},
\end{equation}

\begin{equation}
\tilde{D}_{*}y_{j}(t)=\mathbf{b}_{j*}.
\end{equation}
Accordingly, the continuity equation (7) becomes 

\begin{equation}
\frac{\partial\rho}{\partial t}=-\frac{\partial}{\partial x_{cm}}\left(\rho v_{cm}\right)-\sum_{j=2}^{N}\frac{\partial}{\partial y_{j}}\left[\rho v_{j}\right],
\end{equation}
and the forward stochastic differential equations of motion for $x_{cm}(t)$
and $y_{j}(t)$, respectively, take the form

\begin{equation}
dx_{cm}(t)=\left[\frac{1}{M}\frac{\partial S(x_{cm},\mathbf{y}(t),t)}{\partial x_{cm}}+\frac{\hbar}{2M}\frac{1}{\rho(x_{cm},\mathbf{y}(t),t)}\frac{\partial\rho(x_{cm},\mathbf{y}(t),t)}{\partial x_{cm}}\right]|_{x_{cm}=x_{cm}(t)}dt+dW_{cm}(t),
\end{equation}
and

\begin{equation}
dy_{j}(t)=\left[\frac{1}{m}\frac{\partial S(x_{cm}(t),\mathbf{y},t)}{\partial y_{j}}+\frac{\hbar}{2m}\frac{1}{\rho(x_{cm}(t),\mathbf{y},t)}\frac{\partial\rho(x_{cm}(t),\mathbf{y},t)}{\partial y_{j}}\right]|_{\mathbf{y}=\mathbf{y}(t)}dt+dW_{j}(t),
\end{equation}
The $dW_{cm}$ and $dW_{j}$ are 1-dimensional Wiener processes satisfying
Gaussianity, independence of $dx_{cm}(s)$ and $dy_{j}(s)$ for $s\leq t$,
and variances 
\begin{equation}
E_{t}\left[dW_{cm}^{2}\right]=\frac{\hbar}{M}dt,
\end{equation}
\begin{equation}
E_{t}\left[dW_{j}^{2}\right]=\frac{\hbar}{m}dt,
\end{equation}
respectively. Analogous relations for the backward stochastic differential
equations can be written down as well.

We emphasize that (14) is equivalent to (1), the two being related
by the coordinate transformations (12-13). Thus, applying 
\begin{equation}
J(x_{cm},\mathbf{y},t)=extremal,
\end{equation}
we obtain 

\begin{equation}
\frac{M}{2}\left[\tilde{D}_{*}\tilde{D}+\tilde{D}\tilde{D}_{*}\right]x_{cm}(t)+\frac{m}{2}\sum_{j=2}^{N}\left[\tilde{D}_{*}\tilde{D}+\tilde{D}\tilde{D}_{*}\right]y_{j}(t)=-\left[\frac{\partial}{\partial x_{cm}}U(X,t)|_{X=X(t)}+\sum_{j=2}^{N}\frac{\partial}{\partial y_{j}}U(X,t)|_{X=X(t)}\right].
\end{equation}
By D'Alembert's principle, the variations $\delta x_{cm}(t)$ and
$\delta\mathbf{y}(t)$ are independent of each other, and the $\delta y_{j}(t)$
are independent for all $j$. So (29) separates into the pair

\begin{equation}
\frac{M}{2}\left[\tilde{D}_{*}\tilde{D}+\tilde{D}\tilde{D}_{*}\right]x_{cm}(t)=-\frac{\partial}{\partial x_{cm}}U(X,t)|_{X=X(t)}=-\sum_{i=1}^{N}\frac{\partial}{\partial x_{i}}U_{ext}(x_{i})|_{x_{i}=x_{i}(t)},
\end{equation}

\begin{equation}
\frac{m}{2}\sum_{j=2}^{N}\left[\tilde{D}_{*}\tilde{D}+\tilde{D}\tilde{D}_{*}\right]y_{j}(t)=-\sum_{j=2}^{N}\frac{\partial}{\partial y_{j}}U(X,t)|_{X=X(t)}=-\frac{1}{2}\sum_{j=2}^{N}\sum_{j,k=1}^{N(j\neq k)}\frac{\partial}{\partial x_{j}}U_{int}(x_{j}-x_{k})|_{X=X(t)},
\end{equation}
and (31) separates into 
\begin{equation}
\frac{m}{2}\left[\tilde{D}_{*}\tilde{D}+\tilde{D}\tilde{D}_{*}\right]y_{j}(t)=-\frac{\partial}{\partial y_{j}}U(X,t)|_{X=X(t)},
\end{equation}
for all $j$ from $2,...,N$. The last equality on the right hand
side (rhs) of (30) follows from the fact that the symmetry of $U_{int}$
implies no net force on the CM, and the observation that $\partial x_{i}/\partial x_{cm}=1$
which follows from inverting (12); the last equality on the rhs of
(31-32) follows from the fact that the force on the relative degrees
of freedom come only from $U_{int}$. 

Computing the derivatives on the left sides of (30-31), and removing
the evaluation at $X=X(t)$ on both sides, we obtain

\begin{equation}
M\left[\partial_{t}\mathbf{v}_{cm}+\mathbf{v}_{cm}\frac{\partial}{\partial x_{cm}}\mathbf{v}_{cm}-\mathbf{u}_{cm}\frac{\partial}{\partial x_{cm}}\mathbf{u}_{cm}-\frac{\hbar}{2M}\frac{\partial^{2}}{\partial x_{cm}^{2}}\mathbf{u}_{cm}+\sum_{j=2}^{N}\left(\mathbf{v}_{j}\frac{\partial}{\partial y_{j}}\mathbf{v}_{cm}-\mathbf{u}_{j}\frac{\partial}{\partial y_{j}}\mathbf{u}_{cm}-\frac{\hbar}{2m}\frac{\partial^{2}}{\partial y_{j}^{2}}\mathbf{u}_{cm}\right)\right]=-\frac{\partial}{\partial x_{cm}}U,
\end{equation}

\begin{equation}
m\sum_{j=2}^{N}\left[\partial_{t}\mathbf{v}_{j}+\sum_{j=2}^{N}\left(\mathbf{v}_{j}\frac{\partial}{\partial y_{j}}\mathbf{v}_{j}-\mathbf{u}_{j}\frac{\partial}{\partial y_{j}}\mathbf{u}_{j}\right)-\frac{\hbar}{2M}\frac{\partial^{2}}{\partial x_{cm}^{2}}\mathbf{u}_{j}-\frac{\hbar}{2m}\sum_{j=2}^{N}\frac{\partial^{2}}{\partial y_{j}^{2}}\mathbf{u}_{j}+\mathbf{v}_{cm}\frac{\partial}{\partial x_{cm}}\mathbf{v}_{j}-\mathbf{u}_{cm}\frac{\partial}{\partial x_{cm}}\mathbf{u}_{j}\right]=-\sum_{j=2}^{N}\frac{\partial}{\partial y_{j}}U,
\end{equation}
where $\mathbf{v}_{cm}$ ($\mathbf{v}_{j}$) and $\mathbf{u}_{cm}$
($\mathbf{u}_{j}$) are now velocity fields over Gibbsian ensembles
of (CM and relative) particles. Thus, by integrating the positional
derivatives on both sides of (33) and (34), respectively, and setting
the arbitrary integration constants equal to zero (for simplicity),
each equation yields the quantum Hamilton-Jacobi equation in CM and
relative coordinates: 

\begin{equation}
-\partial_{t}S=U+\frac{1}{2M}\left(\frac{\partial S}{\partial x_{cm}}\right)^{2}-\frac{\hbar^{2}}{2M}\frac{1}{\sqrt{\rho}}\frac{\partial^{2}}{\partial x_{cm}^{2}}\sqrt{\rho}+\sum_{j=2}^{N}\left[\frac{1}{2m}\left(\frac{\partial S}{\partial y_{j}}\right)^{2}-\frac{\hbar^{2}}{2m}\frac{1}{\sqrt{\rho}}\frac{\partial^{2}}{\partial y_{j}^{2}}\sqrt{\rho}\right].
\end{equation}
Combining with (35) with (23), the Madelung transformation yields
the coordinate-transformed Schr{\"o}dinger equation 

\begin{equation}
i\hbar\frac{\partial\psi}{\partial t}=\left[-\frac{\hbar^{2}}{2M}\frac{\partial^{2}}{\partial x_{cm}^{2}}-\frac{\hbar^{2}}{2m}\sum_{j=2}^{N}\frac{\partial^{2}}{\partial y_{j}^{2}}+U\right]\psi,
\end{equation}
where $\psi(x_{cm},\mathbf{y},t)=\sqrt{\rho(x_{cm},\mathbf{y},t)}e^{iS(x_{cm},\mathbf{y},t)/\hbar}$
is single-valued and smooth (because we're restricted to the configuration
space of dimension $\mathbb{R}^{N}$). As shown by Oriols et al. \cite{Oriols2016},
(36) corresponds to the N-particle Schr{\"o}dinger equation 

\begin{equation}
i\hbar\frac{\partial\psi}{\partial t}=\left[-\frac{\hbar^{2}}{2m}\sum_{i=1}^{N}\frac{\partial^{2}}{\partial x_{i}^{2}}+U\right]\psi,
\end{equation}
where $\psi(X,t)=\sqrt{\rho(X,t)}e^{iS(X,t)/\hbar}$, under the coordinate
transformations (12-13). 

From the solution of (36), we can rewrite (24-25) as

\begin{equation}
dx_{cm}(t)=\left[\frac{\hbar}{M}\mathrm{Im}\frac{\partial}{\partial x_{cm}}\ln\psi(x_{cm},\mathbf{y}(t),t)+\frac{\hbar}{M}\mathrm{Re}\frac{\partial}{\partial x_{cm}}\ln\psi(x_{cm},\mathbf{y}(t),t)\right]|_{x_{cm}=x_{cm}(t)}dt+dW_{cm}(t),
\end{equation}

\begin{equation}
dy_{j}(t)=\left[\frac{\hbar}{m}\mathrm{Im}\frac{\partial}{\partial y_{j}}\ln\psi(x_{cm}(t),\mathbf{y},t)+\frac{\hbar}{m}\mathrm{Re}\frac{\partial}{\partial y_{j}}\ln\psi(x_{cm}(t),\mathbf{y},t)\right]|_{\mathbf{y}=\mathbf{y}(t)}dt+dW_{j}(t).
\end{equation}
Given an initial wavefunction $\psi(x_{cm},\mathbf{y},0)$ and initial
trajectories $\left\{ x_{cm}^{h}(0),\mathbf{y}^{h}(0)\right\} $,
where the $h$ index labels a particular set of possible initial trajectories,
the stochastic kinematical evolution of the CM and relative coordinates
can be determined completely. 

Let's now consider the 2nd-order time-evolution of the mean trajectories
of the CM. Defining 
\begin{equation}
Q_{cm}\coloneqq-\frac{\hbar^{2}}{2M}\frac{1}{\sqrt{\rho}}\frac{\partial^{2}}{\partial x_{cm}^{2}}\sqrt{\rho},
\end{equation}
\begin{equation}
\sum_{j=2}^{N}Q_{j}\coloneqq-\frac{\hbar^{2}}{2m}\sum_{j=2}^{N}\frac{1}{\sqrt{\rho}}\frac{\partial^{2}}{\partial y_{j}^{2}}\sqrt{\rho},
\end{equation}
we can rewrite (33) as 

\begin{equation}
M\frac{d^{2}x_{cm}(t)}{dt^{2}}=M\left[\partial_{t}\mathbf{v}_{cm}+\mathbf{v}_{cm}\frac{\partial}{\partial x_{cm}}\mathbf{v}_{cm}+\sum_{j=2}^{N}\mathbf{v}_{j}\frac{\partial}{\partial y_{j}}\mathbf{v}_{cm}\right]|_{x_{cm}=x_{cm}(t)}^{\mathbf{y}=\mathbf{y}(t)}=-\frac{\partial}{\partial x_{cm}}\left[U+Q_{cm}+\sum_{j=2}^{N}Q_{j}\right]|_{x_{cm}=x_{cm}(t)}^{\mathbf{y}=\mathbf{y}(t)},
\end{equation}
and the $j$-th component of (34) as

\begin{equation}
m\frac{d^{2}y_{j}(t)}{dt^{2}}=m\left[\partial_{t}\mathbf{v}_{j}+\sum_{j=2}^{N}\mathbf{v}_{j}\frac{\partial}{\partial y_{j}}\mathbf{v}_{j}+\mathbf{v}_{cm}\frac{\partial}{\partial x_{cm}}\mathbf{v}_{j}\right]|_{x_{cm}=x_{cm}(t)}^{\mathbf{y}=\mathbf{y}(t)}=-\frac{\partial}{\partial y_{j}}\left[U+Q_{cm}+\sum_{j=2}^{N}Q_{j}\right]|_{x_{cm}=x_{cm}(t)}^{\mathbf{y}=\mathbf{y}(t)}.
\end{equation}
We note that equation (42) corresponds to equation (14) of Oriols
et al. \cite{Oriols2016}. 

Now, consider N experimental preparations \footnote{`Experimental' could refer to an actual laboratory experiment or a
natural physical process outside laboratories.} of a system of N identical particles, described by (36), each with
the same initial wavefunction $\psi(x_{cm},\mathbf{y},0)$. For each
preparation, there will be a different set of ``$h$-trajectories''
\cite{Oriols2016}, and because of the identicality of the particles,
they will all have the same marginal probability distribution:
\begin{equation}
\bar{\rho}(y_{k},0)\coloneqq\frac{1}{M}\sum_{h=1}^{M}\delta(y-y_{k}^{h}(0)),
\end{equation}
where M is a very large number of preparations. When $N\rightarrow\infty,$
the distribution of initial particle positions in a single $h$-preparation,
\begin{equation}
P(y_{k},0)\coloneqq\frac{1}{N}\sum_{h=1}^{N}\delta(y-y_{k}^{h}(0)),
\end{equation}
fills the entire support of (44), thereby giving 
\begin{equation}
\bar{\rho}(y_{k},0)\approx P(y_{k},0)
\end{equation}
for the vast majority of the N preparations, where `vast majority'
refers to the possible set of N initial trajectories $X^{h}(t)=\left\{ x_{1}^{h}(t),...,x_{N}^{h}(t)\right\} $
selected according to the initial probability density $\rho(X,0)=|\psi(X,0)|^{2}$.
The fact that the possible set of initial trajectories is selected
randomly according to $|\psi(X,0)|^{2}$ ensures that possible sets
of initial trajectories which don't satisfy (46) will be extremely
rare; and because the $|\psi(X,0)|^{2}$ distribution is preserved
in time by the equivariant evolution given by (23), such possible
sets of initial trajectories not satisfying (46) will be extremely
rare for all times. Thus, Oriols et al. \cite{Oriols2016} refer to
wavefunctions with probability densities satisfying (46) as ``wavefunctions
full of particles'' (WFPs). 

As noted by Oriols et al. \cite{Oriols2016}, however, there are N-particle
wavefunctions which don't satisfy (46). For example, a factorizable
wavefunction $\psi(X,0)=\prod_{i=1}^{N}\phi(x_{i},0)$ in general
won't be a WFP because it won't have the necessary bosonic or fermionic
symmetry requirements to justify the independence of the marginal
probability distributions for each $x_{i}$ (the only exception being
a bosonic wavefunction where all the $\phi_{i}$ are equal, such as
in the mean-field approximation). For another example, wavefunctions
with strong quantum correlations between the particles (see equation
D3 of Oriols et al. \cite{Oriols2016} for an example involving an
unphysical macroscopic superposition state) won't have a single $h$-preparation
which, in the limit $N\rightarrow\infty$, fills the entire support
of (44); however, Oriols et al. \cite{Oriols2016} argue that ``most
of the wave functions associated to macroscopic objects fulfill the
requirements of a wave function full of particles, i.e. they do not
include strong quantum correlations between particles'' (page 12).
While they don't explain why they argue that most wavefunctions associated
to macroscopic objects don't include strong quantum correlations between
particles, their expectation can be justified from the following observation:
in dBB and stochastic mechanics, macroscopic superposition states
(in the real world) arise as a result of decoherence from system-environment
interactions \cite{Goldstein1987,Jibu1990,Blanchard1992,Peruzzi1996,Duerr2009,Goldstein2013},
and such decoherence is always accompanied by ``effective collapse''
\cite{Goldstein1987,Jibu1990,Blanchard1992,Peruzzi1996,Duerr2009,Goldstein2013}.
Effective collapse being the process whereby a dBB/Nelsonian/ZSM particle
(or collection of such particles) composing the system dynamically
evolves into one of the effective system wavefunction components of
a system-environment entangled state, the latter formed during an
environmental decoherence process. Thus effective collapse ensures
that, for all practical purposes, only \emph{one} of the components
of a system-environment entangled state (i.e., a macroscopic quantum
superposition state) will be dynamically relevant to the future motion
of a single-particle or multi-particle system coupled to a macroscopic
environment. In other words, an environmentally decohered macroscopic
object (composed of dBB/Nelsonian/ZSM particles), which are virtually
all of the macroscopic objects in the real world (according to dBB
and stochastic mechanics), can always be expected to have a many-particle
effective wavefunction associated to it corresponding to a WFP. In
section 6, we will say more about how environmental decoherence and
effective collapse of large N systems might be modeled within the
Oriols et al. scheme. In the mean time, we will continue with assuming
pure states that satisfy (46) and thus correspond to WFPs.

Focusing now on the CM motion given by (42) and (38), we shall specify
the conditions under which its classical limit is obtained. For convenience,
we rewrite (42) as 

\begin{equation}
M\frac{d^{2}\mathbf{x}_{cm}(t)}{dt^{2}}=F_{U}+F_{cm}+F_{rel}.
\end{equation}
Classicality conditions will be obtained from comparing the N-dependences
of the three forces on the rhs of (47). 

First we recall that, because of the symmetry of $U_{int}$, its net
force on the CM is zero, leaving the only non-zero net force coming
from $U_{ext}$: 
\begin{equation}
F_{U}\coloneqq-\frac{\partial}{\partial x_{cm}}\sum_{i=1}^{N}U_{ext}(x_{i})=-\sum_{i=1}^{N}\frac{\partial U_{ext}(x_{i})}{\partial x_{i}}.
\end{equation}
Furthermore, if spatial variations of $U_{ext}$ are much larger than
the size of the N-particle system under consideration \footnote{More precisely, if, for an N-particle system with number density of
width $d$, in the presence of an external potential $U$ with scale
of spatial variation given by $L(U)=\sqrt{|\frac{U'}{U'''}|}$, we
have that $d\ll L(U)$. This statement is closely related to the classicality
condition used in the Ehrenfest theorem and in the quantum-classical
limit scheme of Allori et al. \cite{Allori2001,Allori2002,Duerr2009},
i.e., that the de Broglie wavelength $\lambda$ of a single-particle
wavepacket of width $\sigma$ (where $\sigma\geq\lambda$) satisfies
$\lambda\ll L(U)$.} (which will typically be the case for classical external potentials
on macroscopic lengthscales), then (48) can be approximated as (using
$\partial x_{i}/\partial x_{cm}=1$) 
\begin{equation}
F_{U}=F_{ext}\approx-N\frac{\partial U_{ext}(x_{cm})}{\partial x_{cm}},
\end{equation}
which is exact for linear and quadratic potentials as pointed out
by Oriols et al. \cite{Oriols2016}. Thus we have that $F_{U}\propto N$. 

Second, Oriols et al. \cite{Oriols2016} note that the conditional
probability distribution for the CM position can be found by considering
the probability distribution of $x_{cm}^{h}(t)$ for a large number
of different $h$-trajectories given by $X^{h}(t)=\left\{ x_{1}^{h}(t),...,x_{N}^{h}(t)\right\} $.
For a WFP in the limit $N\rightarrow\infty$, the second and third
moments of the distribution are zero (see Theorems 9-10 of Appendix
D of Oriols et al. \cite{Oriols2016}). Hence, for very large but
finite N, one can expect a normal distribution for the CM position:
\begin{equation}
\rho(x_{cm}^{h}(t))\approx\frac{1}{\sqrt{2\pi}\sigma_{cm}}exp\left(-\frac{\left[\bar{x}-x_{cm}^{h}(t)\right]^{2}}{2\sigma_{cm}^{2}}\right),
\end{equation}
where $\sigma_{cm}$ is estimated (see Theorem 10 in Appendix D of
Oriols et al. \cite{Oriols2016}) to be given by 
\begin{equation}
\sigma_{cm}^{2}\leq\frac{\sigma^{2}}{N},
\end{equation}
where $\sigma^{2}$ is the variance of the marginal distributions,
and where $\bar{x}$ is the mean position of the CM. The relation
between these last two variables can be seen as follows. First, we
have (from Corollary 1 and Theorem 4 in Appendix B of Oriols et al.
\cite{Oriols2016}) that 
\begin{equation}
\bar{x}\equiv\bar{x}_{i}=\int_{-\infty}^{\infty}dx\,x\,\bar{\rho}(x),
\end{equation}
where $\bar{\rho}(x)$ is the marginal probability density of the
\emph{i}-th particle \footnote{This is defined as $\bar{\rho}(x)\equiv\bar{\rho}_{i}(x_{i})\coloneqq\int_{-\infty}^{\infty}dx{}_{1}...\int_{-\infty}^{\infty}dx{}_{i-1}\int_{-\infty}^{\infty}dx_{i+1}...\int_{-\infty}^{\infty}dx_{N}\rho(X)$.
Furthermore, for identical particles, the marginal probability density
satisfies $\bar{\rho}_{i}(x_{i})=\bar{\rho}_{j}(x_{j})$ for $i\neq j$
(see the proof of Theorem 3 in Appendix B of Oriols et al. \cite{Oriols2016}).} from which it follows (from Corollary 2 in Appendix B of Oriols et
al. \cite{Oriols2016}) that
\begin{equation}
\sigma^{2}\equiv\sigma_{i}^{2}=\int_{-\infty}^{\infty}dx\,\left(x-\bar{x}\right)\,\bar{\rho}(x).
\end{equation}
Now, calculating $Q_{cm}$ and $F_{cm}$ in terms of $\rho(x_{cm}^{h}(t))$,
one obtains
\begin{equation}
Q_{cm}\approx\frac{\hbar}{2M\sigma_{cm}^{2}}\left(1-\frac{\left[\bar{x}-x_{cm}(t)\right]^{2}}{\sigma_{cm}^{2}}\right),
\end{equation}
\begin{equation}
F_{cm}\approx-\frac{\partial Q_{cm}}{\partial x_{cm}}|_{x_{cm}=x_{cm}(t)}^{\mathbf{y}=\mathbf{y}(t)}\propto\frac{\hbar}{m\sigma^{3}}\sqrt{N},
\end{equation}
where it is used that $\bar{x}-x_{cm}(t)\approx\sigma/\sqrt{N}$.
Thus $F_{cm}\propto\sqrt{N}$. 

Third, since we are dealing with identical particles, we have that
$\rho(x_{cm},y_{2},...,y_{j},...)=\rho(x_{cm},y_{j},...,y_{2},...)$,
and so the force $F_{rel}$ can be rewritten as
\begin{equation}
F_{rel}\coloneqq\frac{\hbar^{2}}{2m}\sum_{j=2}^{N}\left[\frac{\partial}{\partial x_{cm}}\left(\frac{1}{\sqrt{\rho}}\frac{\partial^{2}\sqrt{\rho}}{\partial y_{2}^{2}}\right)\right]|_{x_{cm}=x_{cm}(t)}^{\mathbf{y}=\mathbf{y}(t)}.
\end{equation}
As emphasized by Oriols et al., the exchange symmetry in $\rho$ means
that a single preparation with $N\rightarrow\infty$ is equivalent
to $h=\left\{ 1,...,N\right\} $ different preparations with $y_{2}^{h}(t)$
approximately filling the entire support of $\rho$ in the $y_{2}$
3-space. Accordingly, the quantum equilibrium distribution for the
particles implies that the sum in (56) can be approximated by an integral
that's weighted by $\rho$: 
\begin{equation}
F_{rel}\approx N\frac{\hbar^{2}}{2m}\int_{y_{2}}\left[\rho\frac{\partial}{\partial x_{cm}}\left(\frac{1}{\sqrt{\rho}}\frac{\partial^{2}\sqrt{\rho}}{\partial y_{2}^{2}}\right)\right]|_{x_{cm}=x_{cm}(t)}^{y_{3}(t),...,y_{N}(t)}dy_{2}\rightarrow0.
\end{equation}
That (57) vanishes is due to a symmetric distribution of positive
and negative summands (for an explicit proof, see Appendix E of Oriols
et al. \cite{Oriols2016}). Thus $F_{rel}\approx0$ as $N\rightarrow\infty$.

To summarize, then, in the limit that $N\rightarrow\infty$ for identical
particles, we have 
\begin{equation}
\begin{array}{c}
F_{U}\propto N,\\
\\
F_{cm}\propto\sqrt{N},\\
\\
F_{rel}\rightarrow0.
\end{array}
\end{equation}
So it is clear that the classical external force $F_{U}$ grows much
faster (under the stated conditions) than the two quantum forces,
as the number of identical particles interacting through $U_{int}$
becomes very large. This conclusion does not hold, of course, for
the relative degrees of freedom, but since the CM motion is the only
one that's relevant on macroscopic lengthscales, this poses no problem.
An interesting consequence of (58) is that quantum uncertainty becomes
negligible: between any two preparations of an N-particle system,
$X^{h}(t)$ and $X^{l}(t)$, the CM trajectories and velocities will
be very similar, i.e. $x_{cm}^{h}(t)\approx x_{cm}^{l}(t)$ and $\mathbf{v}_{cm}^{h}(t)\approx\mathbf{v}_{cm}^{l}(t)$.

How large does N have to be for $F_{cm}$ and $F_{rel}$ to become
negligible relative to $F_{U}$? This was addressed by Oriols et al.
in numerical simulations \cite{Oriols2016}. 

In one simulation (Appendix F of Oriols et al. \cite{Oriols2016}),
an initial N-particle wavefunction for identical particles was constructed
from pairs of Gaussian wave packets, with random dispersion and opposite
random momenta and central positions (in other words, the packets
move towards each other and eventually interfere), under the action
of an external linear potential. The linear potential spans a lengthscale
of $\sim10^{-7}m$, while the packet widths are only $\sim10^{-10}m$,
thereby satisfying the condition that the classical external potential
varies over lengthscales much greater than the size of the N-particle
system. Half of the initial positions of the N particles were selected
randomly according to the probability density of the left packet,
the other half according to the probability density of the right packet,
and then the evolution of the CM was computed under the influence
of the three forces in (47). As a comparison, the classical CM was
computed from Newton's law with the linear potential (i.e., $F_{U}$
alone), with the same initial CM position and velocity. The resulting
trajectories were compared for N = 1 through N = 20 (see Figure 1
of Oriols et al. \cite{Oriols2016}). For N = 1, the relative error
between the classical and quantum CM motions increases from zero to
45\% in 2 picoseconds; for N = 20, the relative error increase drops
to less than 2\% in the same duration. In other words, for N = 1,
the classical and quantum CM motions significantly differ from each
other in a very short time, as expected, while for N = 20, the two
CM motions become effectively indistinguishable in a very short time.
Moreover, even for N distinguishable particles, under the same conditions,
Oriols et al. find that the relative error for N = 20 decreases to
around 5\% in 2 picoseconds. It is remarkable that, under the stated
conditions, relatively few particles are needed to reach the ``large
N'' regime. 

As a corollary to the above results, we note that, for the case of
a WFP, the CM osmotic velocity is given by 
\begin{equation}
\mathbf{u}_{cm}=\frac{\hbar}{2M}\frac{1}{\rho}\frac{\partial\rho}{\partial x_{cm}}|_{x_{cm}=x_{cm}(t)}\approx\frac{\hbar}{2m\sigma\sqrt{N}},
\end{equation}
while from (47) and (49) the CM current velocity is found to be 
\begin{equation}
\mathbf{v}_{cm}=\frac{1}{M}\frac{\partial S}{\partial x_{cm}}|_{x_{cm}=x_{cm}(t)}\approx-\frac{1}{Nm}\int_{t_{0}}^{t}\left[N\frac{\partial U_{ext}}{\partial x_{cm}}-F_{cm}\right]|_{x_{cm}=x_{cm}(t)}dt'+\mathbf{v}_{cm0},
\end{equation}
where the contribution from $F_{rel}$ is neglected because, as we
saw from (57), it rapidly approaches zero in the large N limit. Since
$F_{cm}$ is the only N-dependent term in (60) and scales like $\sqrt{N}$,
we can see that in the large N limit, the dominant contribution to
the CM current velocity will come from $\partial U_{ext}/\partial x_{cm}$.
Accordingly, in the large N limit, the CM current velocity will dominate
the kinematics over the CM osmotic velocity (59). Thus, recalling
the forward stochastic differential equation

\begin{equation}
dx_{cm}(t)=\left[\frac{1}{M}\frac{\partial S}{\partial x_{cm}}+\frac{\hbar}{2M}\frac{1}{\rho}\frac{\partial\rho}{\partial x_{cm}}\right]|_{x_{cm}=x_{cm}(t)}dt+dW_{cm}(t),
\end{equation}
where $E_{t}\left[dW_{cm}^{2}\right]=\frac{\hbar}{M}dt$, we can see
that as $N\rightarrow\infty$, $E_{t}\left[dW_{cm}^{2}\right]\rightarrow0$
and (61) reduces to 
\begin{equation}
\frac{dx_{cm}(t)}{dt}\approx\frac{1}{M}\frac{\partial S_{cl}}{\partial x_{cm}}|_{x_{cm}=x_{cm}(t)}.
\end{equation}
The same follows, of course, for the backward stochastic differential
equation.

Extending the above approach to the case of 3-space is formally straightforward,
and entails the replacements $\partial/\partial x_{cm}\rightarrow\nabla_{cm}$,
$\partial/\partial y_{j}\rightarrow\nabla_{j}$, $x_{cm}\rightarrow\mathbf{R}_{cm}$,
$\mathbf{y}\rightarrow\mathbf{r}$, and inclusion of the quantization
relation for the phase field $S$:

\begin{equation}
\oint_{L}\nabla_{cm}S(\mathbf{R}_{cm},\mathbf{r},t)\cdot d\mathbf{R}_{cm}+\sum_{j=1}^{N-1}\oint_{L}\nabla_{j}S(\mathbf{R}_{cm},\mathbf{r},t)\cdot d\mathbf{r}_{j}=nh.
\end{equation}
This last ensures that the 3N-dimensional generalizations of (23)
and (35) are indeed equivalent to the 3N-dimensional generalization
of (36), and that $\psi(\mathbf{R}_{cm},\mathbf{r},t)$ is single-valued
with (generally) multi-valued phase.

\section{Classical nonlinear Schr{\"o}dinger equation for large N center-of-mass
motion}

\subsection{Oriols et al.'s derivation}

What form does the time-dependent Schr{\"o}dinger equation for $\psi(x_{cm},\mathbf{y},t)$
take in the large N limit? Before presenting our answer, let us first
review and critique the answer given by Oriols et al. \cite{Oriols2016}.

Introduce the conditional $S$ and $\rho$ functions for the CM by
the following definitions:
\begin{equation}
\begin{array}{ccc}
S_{cm}(x_{cm},t)\coloneqq S(x_{cm},\mathbf{y}(t),t), &  & \rho_{cm}(x_{cm},t)\coloneqq\rho(x_{cm},\mathbf{y}(t),t),\end{array}
\end{equation}
where $S_{cm}$ satisfies 
\begin{equation}
-\partial_{t}S_{cm}=\frac{1}{2M}\left(\frac{\partial S_{cm}}{\partial x_{cm}}\right)^{2}+U(x_{cm},\mathbf{y}(t),t)+A,
\end{equation}
with 
\begin{equation}
A\coloneqq Q_{cm}+\sum_{j=2}^{N}\left[\frac{1}{2m}\left(\frac{\partial S_{cm}}{\partial y_{j}}\right)^{2}+Q_{j}-v_{j}^{h}(t)\frac{\partial S_{cm}}{\partial y_{j}}\right],
\end{equation}
and where $\rho_{cm}$ satisfies 
\begin{equation}
-\partial_{t}\rho_{cm}=\frac{\partial}{\partial x_{cm}}\left(\frac{1}{M}\frac{\partial S_{cm}}{\partial x_{cm}}\rho_{cm}\right)+B,
\end{equation}
with 
\begin{equation}
B\coloneqq-\sum_{j=2}^{N}\left[\frac{\partial\rho_{cm}}{\partial y_{j}}v_{j}^{h}(t)-\frac{\partial}{\partial y_{j}}\left(\frac{1}{m}\frac{\partial S_{cm}}{\partial y_{j}}\rho_{cm}\right)\right].
\end{equation}
Using the Madelung transformation, (65) and (67) can then be combined
into the `conditional Schr{\"o}dinger equation' \cite{Derakhshani2016b}
\begin{equation}
i\hbar\partial_{t}\psi_{cm}=-\frac{\hbar^{2}}{2M}\frac{\partial^{2}\psi_{cm}}{\partial x_{cm}^{2}}-\frac{\hbar^{2}}{2m}\sum_{j=2}^{N}\frac{\partial^{2}\Psi(x_{cm},\mathbf{y},t)}{\partial y_{j}^{2}}|_{\mathbf{y}=\mathbf{y}^{h}(t)}+i\hbar\sum_{j=2}^{N}v_{j}^{h}(t)\frac{\partial\Psi(x_{cm},\mathbf{y},t)}{\partial y_{j}}|_{\mathbf{y}=\mathbf{y}^{h}(t)}+U(x_{cm},\mathbf{y}(t),t)\psi,
\end{equation}
where $\psi_{cm}(x_{cm},t)=\sqrt{\rho_{cm}(x_{cm},t)}e^{iS_{cm}(x_{cm},t)/\hbar}$
is the `conditional wavefunction' in polar form.

Now, from the earlier observation that the large N limit implies 
\begin{equation}
\frac{\partial V}{\partial x_{cm}}|_{x_{cm}^{h}=x_{cm}^{h}(t)}\gg\frac{\partial\left(Q_{cm}+\sum_{j=2}^{N}Q_{j}\right)}{\partial x_{cm}}|_{x_{cm}=x_{cm}(t)}^{\mathbf{y}=\mathbf{y}(t)},
\end{equation}
and noting that 
\begin{equation}
0=\left[\frac{\partial}{\partial x_{cm}}\left(\frac{1}{2m}\left(\frac{\partial S_{cm}}{\partial y_{j}}\right)^{2}-v_{j}^{h}(t)\frac{\partial S_{cm}}{\partial y_{j}}\right)\right]|_{x_{cm}^{h}=x_{cm}^{h}(t)}^{\mathbf{y}=\mathbf{y}(t)},
\end{equation}
it follows that $A\approx0$ along the CM trajectory. So (65) effectively
corresponds to the classical Hamilton-Jacobi equation for the CM,
in the large N limit. 

Oriols et al. assert that it is reasonable to assume $B=0$, since
this turns (67) into the standard continuity equation for the large
N CM Gaussian density (50). Thus (65) and (67) can be combined via
the Madelung transformation to get the classical nonlinear Schr{\"o}dinger
equation 
\begin{equation}
i\hbar\partial_{t}\psi_{cl}=\left(-\frac{\hbar^{2}}{2M}\frac{\partial^{2}}{\partial x_{cm}^{2}}+U(x_{cm},t)-Q_{cm}\right)\psi_{cl},
\end{equation}
where $\psi_{cl}(x_{cm},t)=\sqrt{\rho_{cl}(x_{cm},t)}e^{iS_{cl}(x_{cm},t)/\hbar}$
is the `classical wavefunction' for the CM. 

The problem with this derivation, in our view, is that no physical
justification is given for why it is reasonable to take $B=0$. The
fact that such an assumption turns (67) into the standard continuity
equation is of course true, but this doesn't constitute an explanation
for \emph{why} it should be true.

\subsection{Conditional Madelung equations}

To demonstrate effective decoupling of the CM and relative coordinates
in the large N limit, it is convenient to focus first on the relative
coordinates.

Consider the conditional Madelung variables $\rho_{rel}(\mathbf{y},t)\coloneqq\rho(x_{cm}(t),\mathbf{y},t)$
and $S_{rel}(\mathbf{y},t)\coloneqq S(x_{cm}(t),\mathbf{y},t)$, with
evolution equations 
\begin{equation}
\begin{split}\partial_{t}\rho_{rel} & =-\sum_{j=2}^{N}\frac{\partial}{\partial y_{j}}\end{split}
\left[\rho_{rel}\frac{\partial S_{rel}}{\partial y_{j}}\frac{1}{m}\right]-\frac{\partial}{\partial x_{cm}}\left[\rho\frac{\partial S}{\partial x_{cm}}\frac{1}{M}\right]|_{x_{cm}(t)}+\left(v_{cm}(t)\frac{\partial\rho}{\partial x_{cm}}\right)|_{x_{cm}(t)},
\end{equation}
\begin{equation}
\begin{split}-\partial_{t}S_{rel} & =\sum_{j=2}^{N}\frac{1}{2m}\left(\frac{\partial S_{rel}}{\partial y_{j}}\right)^{2}+\sum_{j=2}^{N}\left(-\frac{\hbar^{2}}{2m}\frac{1}{\sqrt{\rho_{rel}}}\frac{\partial^{2}\sqrt{\rho_{rel}}}{\partial y_{j}^{2}}\right)\\
 & +\frac{1}{2M}\left(\frac{\partial S}{\partial x_{cm}}\right)|_{x_{cm}(t)}-\left(\frac{\hbar^{2}}{2M}\frac{1}{\sqrt{\rho}}\frac{\partial^{2}\sqrt{\rho}}{\partial x_{cm}^{2}}\right)|_{x_{cm}(t)}-v_{cm}(t)\frac{\partial S}{\partial x_{cm}}|_{x_{cm}(t)}+U|_{x_{cm}(t)},
\end{split}
\end{equation}
where again 
\begin{equation}
v_{cm}(t)=\frac{dx_{cm}(t)}{dt}=\frac{1}{M}\frac{\partial S}{\partial x_{cm}}|_{x_{cm}(t),\mathbf{y}(t)},
\end{equation}
and 
\begin{equation}
U|_{x_{cm}(t)}=\left[\sum_{j=1}^{N}U_{ext}(x_{j})+\frac{1}{2}\sum_{j,k=1;j\neq k}^{N}U_{int}(x_{j}-x_{k})\right]|_{x_{cm}(t)}=NU_{ext}(x_{cm}(t))+\frac{1}{2}\sum_{j,k=1;j\neq k}^{N}U_{int}(x_{j}-x_{k}).
\end{equation}

We will argue that, in the limit $N\rightarrow\infty$, the `global'
$S$ and $\rho$ variables effectively decouple in $\mathbf{y}$ and
$x_{cm}$, thereby reducing (73-74) to the corresponding effective
Madelung equations for the relative coordinates, and likewise for
the conditional CM Madelung equations.

\subsubsection{Conditional-to-effective continuity equation}

Equation (73) can be rewritten as follows:
\begin{equation}
\begin{split}\partial_{t}\rho_{rel} & =-\sum_{j=2}^{N}\frac{\partial}{\partial y_{j}}\left[\rho_{rel}\frac{\partial S_{rel}}{\partial y_{j}}\frac{1}{m}\right]-\left[\frac{\partial}{\partial x_{cm}}\left(\rho\frac{\partial S}{\partial x_{cm}}\frac{1}{M}\right)-v_{cm}(t)\frac{\partial\rho}{\partial x_{cm}}\right]|_{x_{cm}(t)}\\
 & =-\sum_{j=2}^{N}\frac{\partial}{\partial y_{j}}\left[\rho_{rel}\frac{\partial S_{rel}}{\partial y_{j}}\frac{1}{m}\right]-\left[\frac{1}{M}\frac{\partial S}{\partial x_{cm}}\frac{\partial\rho}{\partial x_{cm}}+\frac{1}{M}\rho\frac{\partial^{2}S}{\partial x_{cm}^{2}}-\frac{1}{M}\frac{\partial S}{\partial x_{cm}}|_{\mathbf{y}(t)}\frac{\partial\rho}{\partial x_{cm}}\right]|_{x_{cm}(t)}.
\end{split}
\end{equation}
We claim that, in the limit $N\rightarrow\infty$, all the terms in
the last bracket on the rhs of (77) contribute only as time-dependent
correction factors, and therefore can be dropped. 

To see this, recall that when $N\rightarrow\infty$ we have 
\begin{equation}
M\frac{d^{2}x_{cm}(t)}{dt^{2}}\approx-\frac{\partial\left(NU_{ext}(x_{cm})\right)}{\partial x_{cm}}|_{x_{cm}(t)}=-N\frac{\partial U_{ext}(x_{cm})}{\partial x_{cm}}|_{x_{cm}(t)},
\end{equation}
where $U_{ext}$ spatially varies on macroscopic scales. Integrating
(78) gives 
\begin{equation}
\frac{dx_{cm}(t)}{dt}=\frac{1}{M}\frac{\partial S(x_{cm},\mathbf{y},t)}{\partial x_{cm}}|_{x_{cm}(t),\mathbf{y}(t)}\approx-\frac{1}{m}\int_{t_{0}}^{t}\left(\frac{\partial U_{ext}}{\partial x_{cm}}\right)|_{x_{cm}(t')}dt'+v_{cm}(0)\eqqcolon\frac{1}{M}\frac{\partial S_{cl}(x_{cm},t)}{\partial x_{cm}}|_{x_{cm}(t)},
\end{equation}
and thus 
\begin{equation}
\frac{1}{M}\frac{\partial^{2}S(x_{cm},\mathbf{y},t)}{\partial x_{cm}^{2}}|_{x_{cm}(t),\mathbf{y}(t)}\approx-\frac{1}{m}\int_{t_{0}}^{t}\left(\frac{\partial^{2}U_{ext}}{\partial x_{cm}^{2}}\right)|_{x_{cm}(t')}dt'\eqqcolon\frac{1}{M}\frac{\partial^{2}S_{cl}(x_{cm},t)}{\partial x_{cm}^{2}}|_{x_{cm}(t)},
\end{equation}
which we see are effectively independent of $\mathbf{y}$ and only
depend on time.

Note, also, that the equation of motion for the relative positions
is given by 
\begin{equation}
m\frac{d^{2}y_{j}(t)}{dt^{2}}=-\frac{\partial}{\partial y_{j}}\left[\frac{1}{2}\sum_{j,k=1;j\neq k}^{N}U_{int}+Q_{cm}+\sum_{j=2}^{N}Q_{j}\right].
\end{equation}
where 
\begin{equation}
\frac{dy_{j}(t)}{dt}=-\frac{1}{m}\int_{t_{0}}^{t}\frac{\partial}{\partial y_{j}}\left[\frac{1}{2}\sum_{j,k=1;j\neq k}^{N}U_{int}+Q_{cm}+\sum_{j=2}^{N}Q_{j}\right]|_{\mathbf{y}(t'),x_{cm}(t')}dt'+v_{j}(0)=\frac{1}{m}\frac{\partial S}{\partial y_{j}}|_{x_{cm}(t),\mathbf{y}(t)}.
\end{equation}
In the limit $N\rightarrow\infty$, we have
\begin{equation}
m\frac{d^{2}y_{j}(t)}{dt^{2}}\approx-\frac{\partial}{\partial y_{j}}\left[\frac{1}{2}\sum_{j,k=1;j\neq k}^{N}U_{int}+\sum_{j=2}^{N}Q_{j}\right]
\end{equation}
and 
\begin{equation}
\frac{dy_{j}(t)}{dt}=\frac{1}{m}\frac{\partial S(x_{cm},\mathbf{y},t)}{\partial y_{j}}|_{x_{cm}(t),\mathbf{y}(t)}\approx-\frac{1}{m}\int_{t_{0}}^{t}\frac{\partial}{\partial y_{j}}\left[\frac{1}{2}\sum_{j,k=1;j\neq k}^{N}U_{int}+\sum_{j=2}^{N}Q_{j}\right]|_{\mathbf{y}(t')}dt'+v_{j}(0)=\frac{1}{m}\frac{\partial S_{rel}(\mathbf{y},t)}{\partial y_{j}}|_{\mathbf{y}(t)},
\end{equation}
since the large N CM density (corresponding to a WFP) takes the form
(50), implying the effective factorization 
\begin{equation}
\underset{N\rightarrow\infty}{lim}\:\rho(x_{cm},\mathbf{y},t)\approx\rho_{cl}(x_{cm},t)\rho_{rel}(\mathbf{y},t),
\end{equation}
which leads to $Q_{cm}$ taking the $\mathbf{y}$-independent form
(54). Correspondingly, for all $j=2,..,N$, equation (85) implies
\begin{equation}
Q_{j}(x_{cm},\mathbf{y},t)\approx-\left(\frac{\hbar^{2}}{2m}\frac{1}{\sqrt{\rho_{rel}(\mathbf{y},t)}}\frac{\partial^{2}\sqrt{\rho_{rel}(\mathbf{y},t)}}{\partial y_{j}^{2}}\right)=Q_{j}(\mathbf{y},t),
\end{equation}
which is effectively independent of $x_{cm}$. 

In other words, in the large N limit, the relative coordinates evolve
in time (effectively) independently of the CM coordinate. 

Accordingly, it follows that (80) only contributes to (77) an uninteresting
time-dependent factor of the form 
\begin{equation}
\frac{1}{M}\left[\rho(x_{cm},\mathbf{y},t)\frac{\partial^{2}S(x_{cm},\mathbf{y},t)}{\partial x_{cm}^{2}}\right]|_{x_{cm}(t)}\approx\frac{1}{M}\left[\rho_{cl}(x_{cm},t)\frac{\partial^{2}S_{cl}(x_{cm},t)}{\partial x_{cm}^{2}}\right]|_{x_{cm}(t)}\rho_{rel}(\mathbf{y},t),
\end{equation}
while (79) along with (85) imply the time-dependent factors 
\begin{equation}
\frac{1}{M}\left(\frac{\partial S}{\partial x_{cm}}\frac{\partial\rho}{\partial x_{cm}}\right)|_{x_{cm}(t)}\approx\frac{1}{M}\left(\frac{\partial S_{cl}(x_{cm},t)}{\partial x_{cm}}\frac{\partial\rho_{cl}(x_{cm},t)}{\partial x_{cm}}\right)|_{x_{cm}(t)}\rho_{rel}(\mathbf{y},t),
\end{equation}
and 
\begin{equation}
\frac{1}{M}\left(\frac{\partial S}{\partial x_{cm}}|_{\mathbf{y}=\mathbf{y}(t)}\frac{\partial\rho}{\partial x_{cm}}\right)|_{x_{cm}(t)}\approx\frac{1}{M}\left(\frac{\partial S_{cl}(x_{cm},t)}{\partial x_{cm}}\frac{\partial\rho_{cl}(x_{cm},t)}{\partial x_{cm}}\right)|_{x_{cm}(t)}\rho_{rel}(\mathbf{y},t).
\end{equation}
Hence, terms (87-89) might as well be dropped from (77), leaving

\begin{equation}
\begin{split}\partial_{t}\rho_{rel} & \approx-\sum_{j=2}^{N}\frac{\partial}{\partial y_{j}}\end{split}
\left[\rho_{rel}\frac{\partial S_{rel}}{\partial y_{j}}\frac{1}{m}\right],
\end{equation}
which is just the effective continuity equation for $\rho_{rel}$.

\subsubsection{Conditional-to-effective quantum Hamilton-Jacobi equation}

Equation (74) can be rewritten as
\begin{equation}
\begin{split}-\partial_{t}S_{rel} & =\sum_{j=2}^{N}\frac{1}{2m}\left(\frac{\partial S_{rel}}{\partial y_{j}}\right)^{2}+\frac{1}{2M}\left(\frac{\partial S}{\partial x_{cm}}\right)^{2}|_{x_{cm}(t)}-\frac{1}{M}\frac{\partial S}{\partial x_{cm}}|_{x_{cm}(t),\mathbf{y}(t)}\cdot\frac{\partial S}{\partial x_{cm}}|_{x_{cm}(t)}\\
 & +\sum_{j=2}^{N}\left(-\frac{\hbar^{2}}{2m}\frac{1}{\sqrt{\rho_{rel}}}\frac{\partial^{2}\sqrt{\rho_{rel}}}{\partial y_{j}^{2}}\right)-\left(\frac{\hbar^{2}}{2M}\frac{1}{\sqrt{\rho}}\frac{\partial^{2}\sqrt{\rho}}{\partial x_{cm}^{2}}\right)|_{x_{cm}(t)}+U|_{x_{cm}(t)}.
\end{split}
\end{equation}
From the arguments in sub-subsection 3.2.1, the large N limit entails
that we can neglect the terms involving $(\partial S/\partial x_{cm})|_{x_{cm}(t)}$,
since they contribute only as time-dependent factors in (91) and drop
out of the equations of motion for the relative coordinates.

Similarly, as noted in subsection 2.1, for large N the center-of-mass
quantum kinetic takes the form (54), which means it contributes only
an uninteresting time-dependent phase shift to $S_{rel}$ in (91).
And, as we showed in subsection 2.1, that the CM quantum kinetic takes
the $\mathbf{y}$-independent form (54) means that the CM quantum
kinetic drops out of the equations of motion for the relative positions,
i.e., equations (81-82), and might as well also be dropped from (91).

Likewise, in $U|_{x_{cm}(t)}$, the external potential component $\sum_{j=1}^{N}U_{ext}(x_{j})=NU_{ext}(x_{cm})$
will also contribute to $S_{rel}$ only a time-dependent phase shift,
and thus can be dropped as well. 

We are thereby left with the effective quantum Hamilton-Jacobi equation
\begin{equation}
\begin{split}-\partial_{t}S_{rel} & \approx\sum_{j=2}^{N}\frac{1}{2m}\left(\frac{\partial S_{rel}}{\partial y_{j}}\right)^{2}+\sum_{j=2}^{N}\left(-\frac{\hbar^{2}}{2m}\frac{1}{\sqrt{\rho_{rel}}}\frac{\partial^{2}\sqrt{\rho_{rel}}}{\partial y_{j}^{2}}\right)+\frac{1}{2}\sum_{j,k=1;j\neq k}^{N}U_{int}.\end{split}
\end{equation}
Accordingly, we conclude that the `global' $S$ function effectively
decomposes as 
\begin{equation}
\underset{N\rightarrow\infty}{lim}\:S(x_{cm},\mathbf{y},t)\approx S_{cl}(x_{cm},t)+S_{rel}(\mathbf{y},t),
\end{equation}
where $S_{cl}(x_{cm},t)$ evolves autonomously by its effective classical
Hamilton-Jacobi equation (equation (65) with $A\approx0$), and likewise
for $\rho_{cl}(x_{cm},t)$ (equation (67) with $B\approx0$). The
Madelung transformation involving $S_{cl}(x_{cm},t)$ and $\rho_{cl}(x_{cm},t)$
then yields the classical nonlinear Schr{\"o}dinger equation (72).

\subsection{Comments on the classical nonlinear Schr{\"o}dinger equation}

As is well known \cite{Schiller1962,Rosen1964,HollandBook1993,Ghose2002,Nikolic2006,Nikolic2007,Oriols2016,Derakhshani2016a,Derakhshani2016b},
(72) can also be formally derived (with $\hbar$ as a free parameter)
from classical statistical mechanics of a single particle in an external
scalar potential, in the Hamilton-Jacobi representation. What's different
here is that (72) is an approximate description of the Schr{\"o}dinger
evolution for the CM of an N-particle system, with potential $U$
(where the external component spatially varies on scales larger than
the size of the N-particle system), in the limit that $N\rightarrow\infty$. 

In order to verify the robustness of (72) as an approximation to classical
dynamics, Oriols et al. \cite{Oriols2016} numerically simulated a
Gaussian wavepacket, defined by taking the square root of (50) and
multiplying by $exp\left(ik_{0}x_{cm}\right)$, evolving by (72) for
two cases: a packet in free fall in external potential $U=2x_{cm}$,
and a packet oscillating in a harmonic oscillator potential $U=x_{cm}^{2}/2$.
In both cases (Figures 2 and 3 of \cite{Oriols2016}), their simulations
confirm that the packets do not disperse over time, and the CM trajectories
(for different initial positions) closely mimic the CM trajectories
one expects from classical mechanics. 

Extending our derivation of (72) to the 3-dimensional case is formally
straightforward, and requires inclusion of the quantization condition
on the 3-dimensional generalization of the CM conditional phase field
as follows:

\begin{equation}
\oint_{L}\mathbf{\nabla}_{cm}S(\mathbf{R}_{cm},\mathbf{r}(t),t)\cdot d\mathbf{R}_{cm}=\oint_{L}\mathbf{\nabla}_{cm}S_{cm}\cdot d\mathbf{R}_{cm}=nh.
\end{equation}
This assures that the 3-dimensional version of $\psi_{cm}$ is single-valued
with (generally) multi-valued phase.

A notable advantage of (72) as a `large N' approximation is that,
in contrast to the mean-field SN and stochastic SN equations, (72)
does not admit macroscopic superpositions of CM position states, and
so does not predict macroscopic semiclassical gravitational/electrostatic
cat states in the case that $U_{int}$ corresponds to an N-body Newtonian
gravitational/Coulomb potential (e.g., such as in a neutron star or
the sun). Basically, this is because the nonlinearity of (72) means
that any pair of solutions, $\psi_{1}^{cl}$ and $\psi_{2}^{cl}$,
cannot be superposed to form a new solution. Thus, only one positional
wavepacket is associated to the evolution of the CM at any time.

\section{Recovering classical Newtonian gravity for many macro particles}

Suppose now that we have K many-particle systems, with CM masses $\left\{ M^{i},...,M^{K}\right\} $,
where the \emph{i}-th CM `particle' is described by a pair of CM Madelung
variables $\left\{ \rho_{cl}^{i},S_{cl}^{i}\right\} $ evolving by
their own effective Madelung equations. Suppose, further, that these
CM particles classically interact via macroscopically long-range classical
gravitational (or electrostatic) potentials (i.e., potentials spatially
varying on scales much larger than the sizes of the N-particle systems
composing the CM particles). Then the K-body effective Madelung equations
for these gravitationally interacting CM particles, are given by
\begin{equation}
-\partial_{t}\rho_{cl}^{K}\approx\sum_{i=1}^{K}\frac{\partial}{\partial x_{cm}^{i}}\left(\frac{1}{M^{i}}\frac{\partial S_{cl}^{K}}{\partial x_{cm}^{i}}\rho_{cl}^{K}\right),
\end{equation}
\begin{equation}
-\partial_{t}S_{cl}^{K}\approx\sum_{i=1}^{K}\frac{1}{2M^{i}}\left(\frac{\partial S_{cl}^{K}}{\partial x_{cm}^{i}}\right)^{2}+\sum_{i=1}^{K}U^{i},
\end{equation}
where the solution of (95) is a product state of narrow Gaussians
\begin{equation}
\rho_{cl}^{K}\approx\prod_{i=1}^{K}\frac{1}{\sqrt{2\pi}\sigma_{cm}^{i}}exp\left(-\frac{\left[\bar{x}^{i}-x_{cm}^{i}\right]^{2}}{2\left[\sigma_{cm}^{i}\right]^{2}}\right)\eqqcolon\prod_{i=1}^{K}\rho_{cl}^{i},
\end{equation}
the solution of (96) takes the form 
\begin{equation}
S_{cl}^{K}\approx\left[\sum_{i=1}^{K}\int p_{cm}^{i}dx_{cm}^{i}-\int\left(\sum_{i=1}^{K}\frac{1}{2M^{i}}\left(p_{cm}^{i}\right)^{2}+\sum_{i=1}^{K}U^{i}\right)dt\right]-\sum_{i=1}^{K}\hbar\phi_{cm}^{i}\eqqcolon\sum_{i=1}^{K}S_{cl}^{i},
\end{equation}
and the potential 
\begin{equation}
U^{i}\coloneqq\sum_{n=1}^{N^{i}}U_{ext}^{i}(x_{n}^{i})+\frac{1}{2}\sum_{j,k=1}^{N^{i}(j\neq k)}U_{int}^{i}(x_{j}-x_{k}),
\end{equation}
where 
\begin{equation}
\sum_{n=1}^{N^{i}}U_{ext}^{i}(x_{n}^{i})\approx N^{i}U_{ext}^{i}(x_{cm}^{i})\coloneqq-\frac{M^{i}}{2}\sum_{l=1}^{K(l\neq i)}\frac{M^{l}}{|x_{cm}^{i}-x_{cm}^{l}|},
\end{equation}
using $(\partial x_{n}^{i}/\partial x_{cm}^{i})=1$.

Notice that, despite the CM `particles' gravitationally interacting
via (100), the large-N CM densities form a product state (97). This
follows from our assumption that the gravitational potentials sourced
by the CM `particles' are (macroscopically) long-range, and therefore
vary on distance scales much larger than the sizes of the N-particle
systems composing the CM particles (i.e., $\sigma_{cm}^{i}\ll\sqrt{|U_{ext}^{i}\prime/U_{ext}^{i}\prime\prime\prime|}$
\cite{Allori2002,Allori2001,Duerr2009}). Recall from subsection 2.1
that this was a necessary condition for showing that the large-N density,
corresponding to a WFP, takes the (approximately) Gaussian form of
the factors in (97). Moreover, although the \emph{zbw} phases of the
CM `particles' are not physically independent, due to the non-separable
potential (100) which physically influences each CM particle via the
(approximately) classical equations of motion

\begin{equation}
M^{i}\frac{dx_{cm}^{i}(t)}{dt}\approx\frac{\partial}{\partial x_{cm}^{i}}S_{cl}^{K}|_{x_{cm}^{i}=x_{cm}^{i}(t)},
\end{equation}

\begin{equation}
M^{i}\frac{d^{2}x_{cm}^{i}(t)}{dt^{2}}\approx-N^{i}\frac{\partial}{\partial x_{cm}^{i}}U_{ext}^{i}(x_{cm}^{i},t)|_{x_{cm}^{i}=x_{cm}^{i}(t)}=M^{i}\frac{\partial}{\partial x_{cm}^{i}}\sum_{l=1}^{K(l\neq i)}\frac{M^{l}}{2|x_{cm}^{i}-x_{cm}^{l}|}|_{x_{cm}^{i}=x_{cm}^{i}(t)}^{x_{cm}^{l}=x_{cm}^{l}(t)},
\end{equation}
it is still meaningful to speak of the \emph{zbw} phase of an individual
CM `particle' in the lab frame; the \emph{i}-th CM `particle', in
the lab frame, has an associated \emph{zbw} phase $S_{cl}^{i}$ that
depends on the sum of all the potentials sourced by the K-1 other
CM `particles', at the space-time location of the \emph{i}-th CM `particle'.
Indeed, the net potential `seen' by an individual CM particle, from
the K-1 CM particles, looks like a slowly varying external potential
as a consequence of $\sigma_{cm}^{i}\ll\sqrt{|U_{ext}^{i}\prime/U_{ext}^{i}\prime\prime\prime|}$.
Thus $S_{cl}^{i}$ varies slowly as a function of $U_{ext}^{i}$ for
all $i=1,...,K$, much like the phase of a light wave moving through
a medium of slowly (spatially) varying refractive index.

If we employ the Madelung transformation, (95-96) can be combined
into the K-body version of (72): 
\begin{equation}
i\hbar\partial_{t}\psi_{cl}^{K}=\sum_{i=1}^{K}\left(-\frac{\hbar^{2}}{2M^{i}}\frac{\partial^{2}}{\partial x_{cm}^{i2}}+U^{i}-Q_{cm}^{i}\right)\psi_{cl}^{K},
\end{equation}
where
\begin{equation}
\psi_{cl}^{K}\approx\prod_{i=1}^{K}\sqrt{\rho_{cl}^{i}}e^{iS_{cl}^{i}/\hbar}\eqqcolon\prod_{i=1}^{K}\psi_{cl}^{i},,
\end{equation}
\begin{equation}
Q_{cm}^{i}\coloneqq-\frac{\hbar^{2}}{2M^{i}}\frac{1}{\sqrt{\rho_{cl}^{K}}}\frac{\partial^{2}}{\partial x_{cm}^{i2}}\sqrt{\rho_{cl}^{K}}\approx\frac{\hbar}{2M^{i}\left(\sigma_{cm}^{i}\right)^{2}}\left(1-\frac{\left[\bar{x}^{i}-x_{cm}^{hi}\right]^{2}}{\left[\sigma_{cm}^{i}\right]^{2}}\right).
\end{equation}
So the Madelung variables for each large-N CM particle define narrow
(in position space) classical wavepackets $\psi_{cl}^{i}$ satisfying
the nonlinear Schr{\"o}dinger equation (103). 

An important property of the K-body system of large-N CM `particles'
is that the CM particle trajectories can cross in configuration space.
To see this, let us recall what the exact dBB/Nelsonian dynamics predict
for a CM `particle' associated to a pure state $\Psi$, when $\Psi$
is a superposition of two (not necessarily narrow) Gaussian wavepackets
in position space moving with fixed speeds in opposite directions
towards each other. When the packets overlap in configuration space,
the exact description says that an ensemble of identical CM particle
trajectories, corresponding to each packet, will not cross but rather
will abruptly (but not discontinuously) change directions and exit
the overlapping region with the packets they did not initially occupy
\cite{HollandBook1993,Bohm1995,Duerr2009}. The physical reason for
this non-classical behavior is that the pure state defines a single-valued
momentum field in configuration space through $p=\hbar\mathrm{Im}\nabla\ln\Psi$,
which means that there will be a unique momentum for each point in
the overlap region. Equivalently, the quantum forces from the quantum
kinetic associated to $\Psi$ in the overlap region push the trajectories
away from each other and causes them to abruptly change directions.
In the case of the K-body system, a superposition of two wavepackets
can't be applied since the packets associated to the large-N CM `particles'
evolve by coupled nonlinear Schr{\"o}dinger equations (103), and any
superposition of two packets doesn't form a new solution of (103).
Nevertheless, we can consider two, identical, large-N CM particles,
associated to two narrow Gaussian wavepackets moving in opposite directions
towards each other and ask if their trajectories will cross (assume
the two particles don't classically interact or only negligibly so).
Yes, because (i) the narrowness of the two packets (recall that $\sigma_{cm}\equiv\frac{\sigma^{2}}{N}$,
and we have $N\rightarrow\infty$, implying that the amplitudes of
the packets are effectively Dirac delta functions) ensures that they
are effectively disjoint (hence don't interfere) in position space,
and (ii) the quantum force is absent from the large N equations of
motion (101-102). So the large-N CM `particles' indeed move like classical
mechanical particles, since classical mechanics predicts that particle
trajectories can cross in configuration space (but not in phase space).
Similar observations have been made by Benseny et al. in \cite{Benseny2016}
and D{\"u}rr et al. in \cite{Duerr2009}.

\section{Recovering classical Vlasov-Poisson mean-field theory}

We can now connect the K-body system of gravitationally interacting,
large-N CM `particles' to the classical Vlasov-Poisson mean-field
theory. 

Assuming the special case of identical CM `particles', multiplying
the first term on the rhs in (99) by $1/K$ (the weak-coupling scaling
CITE), and subtracting out the second term on the rhs in (99) (since
it will only yield a global phase factor), the K-body effective CM
Madelung equations become 
\begin{equation}
-\partial_{t}\rho_{cl}^{K}\approx\sum_{i=1}^{K}\frac{\partial}{\partial x_{cm}^{i}}\left(\frac{1}{M}\frac{\partial S_{cl}^{K}}{\partial x_{cm}^{i}}\rho_{cl}^{K}\right),
\end{equation}
\begin{equation}
H_{cl}^{K}\coloneqq-\partial_{t}S_{cl}^{K}\approx\sum_{i=1}^{K}\frac{1}{2M}\left(\frac{\partial S_{cl}^{K}}{\partial x_{cm}^{i}}\right)^{2}-\frac{M^{2}}{K}\sum_{i=1}^{K}\sum_{l=1}^{K(l\neq i)}\frac{1}{2|x_{cm}^{i}-x_{cm}^{l}|},
\end{equation}
with solutions given by (97-98) for $M^{i}=M$. The classical equations
of motion are just

\begin{equation}
p_{cm}^{i}(t)\coloneqq M\frac{dx_{cm}^{i}(t)}{dt}\approx\frac{\partial}{\partial x_{cm}^{i}}S_{cl}^{K}|_{x_{cm}^{i}=x_{cm}^{i}(t)},
\end{equation}

\begin{equation}
\frac{dp_{cm}^{i}(t)}{dt}=M\frac{d^{2}x_{cm}^{i}(t)}{dt^{2}}\approx\frac{M^{2}}{K}\frac{\partial}{\partial x_{cm}^{i}}\sum_{l=1}^{K(l\neq i)}\frac{1}{2|x_{cm}^{i}-x_{cm}^{l}|}|_{x_{cm}^{i}=x_{cm}^{i}(t)}^{x_{cm}^{l}=x_{cm}^{l}(t)}.
\end{equation}
Now, consider the empirical distribution for the K particles $f_{K}(x_{cm},p_{cm},t)\coloneqq K^{-1}\sum_{i=1}^{K}\delta(x_{cm}-x_{cm}^{i}(t))\delta(p_{cm}-p_{cm}^{i}(t))$
satisfying (in the sense of distributions) the Vlasov equation
\begin{equation}
\begin{split}\partial_{t}f_{K}+p_{cm}\frac{\partial}{\partial x_{cm}}f_{K} & +\frac{\partial}{\partial p_{cm}}\left[F_{K}\left(x_{cm},t\right)f_{K}\right]\\
 & =\frac{1}{K^{2}}\sum_{i=1}^{K}\frac{\partial}{\partial p_{cm}}\left[\left(\frac{\partial}{\partial x_{cm}^{i}}\sum_{l=1}^{K(l\neq i)}\frac{M^{2}}{2|x_{cm}^{i}-x_{cm}^{l}|}|_{x_{cm}^{i}=x_{cm}^{i}(t)}^{x_{cm}^{l}=x_{cm}^{l}(t)}\right)\delta(x_{cm}-x_{cm}^{i}(t))\delta(p_{cm}-p_{cm}^{i}(t))\right],
\end{split}
\end{equation}
 where
\begin{equation}
F_{K}\left(x_{cm},t\right)\coloneqq-\frac{\partial}{\partial x_{cm}}\int_{\mathbb{R}}\int_{\mathbb{R}}\frac{M^{2}}{|x_{cm}-x_{cm}'|}\,f_{K}\,dx_{cm}'dp_{cm}.
\end{equation}

We recall that Golse \cite{Golse2003} and Bardos et al. \cite{BardosGolseMauser2000,BardosErdosGolseMauserYau2002}
considered a D-dimensional generalization of (106-111) \footnote{Regarding (106), Golse \cite{Golse2003} and Bardos et al. \cite{BardosGolseMauser2000,BardosErdosGolseMauserYau2002}
take the empirical position distributions for the particles to be
Dirac delta functions. The 1-dimensional Dirac delta function in position
space is indeed a solution of (106), and note that the factors of
(97) approach 1-dimensional Dirac delta functions as $K\rightarrow\infty$.}, for an arbitrary, symmetric, smooth interaction potential $V$,
and showed that for $K\rightarrow\infty$ one obtains the D-dimensional
Vlasov-Poisson mean-field equations (see also section 4 of Part I).
Thus the system (106-111), in the limit $K\rightarrow\infty$, is
equivalent to the 2-dimensional Vlasov-Poisson mean-field equations
(hereafter, writing $x_{cm}=x$ and $p_{cm}=p$):
\begin{equation}
\partial_{t}f(x,p,t)+\left\{ H^{m.f.}(x,p,t),f(x,p,t)\right\} =0,
\end{equation}

\begin{equation}
H_{cl}^{m.f.}(x,p,t)\coloneqq\frac{p^{2}}{2M}+\int_{\mathbb{R}}M\Phi_{g}^{m.f.}(x,x',t)\,dx'.
\end{equation}

\begin{equation}
\frac{\partial^{2}\Phi_{g}^{m.f.}}{\partial x^{2}}=4\pi M\int_{\mathbb{R}}f(x,p,t)dp=4\pi M\rho(x,t),
\end{equation}
\begin{equation}
F(x,t)\coloneqq-\frac{\partial}{\partial x}\int_{\mathbb{R}}M\Phi_{g}^{m.f.}(x,x',t)\,dx'.
\end{equation}
Extending the above results to the D-dimensional case is formally
straightforward.

\section{Incorporating environmental decoherence}

As discussed in subsection 2.1, environmental decoherence accompanied
by effective collapse ensures that wavefunctions associated to macroscopic
objects (composed of dBB/Nelsonian/ZSM particles) in the real world
will not correspond to macroscopic quantum superpositions (i.e., involve
strong quantum correlations). Thus it is reasonable to expect that
wavefunctions associated to macroscopic objects in the real world
will in general be WFPs. Though this expectation seems reasonable
on general grounds, it would be even more convincing if we could demonstrate
it in an explicit model of environmental decoherence in ZSM (or dBB).
Here we sketch a suggestion for an explicit model. 

There exists a well-known model of generalized Brownian motion in
classical nonequilibrium statistical mechanics called the Kac-Zwanzig
(KZ) model \cite{FordKacMazur1965,Zwanzig1973,EberlingSokolov2005}
(the quantum mechanical analogue is the well-known Caldeira-Leggett
model \cite{CaldeiraLeggett1983a,CaldeiraLeggett1983b}). The KZ model
describes a heavy particle coupled to an external field and a heat
bath, the bath modeled as an n-particle system of light harmonic oscillators,
where the system particle couples bilinearly to each bath particle,
with possibly frequency-dependent coupling strength. The classical
Newtonian equations of motion for the system particle and bath particles
are thereby coupled, and if one integrates out the bath variables,
one finds, under the assumptions that the bath is at thermal equilibrium
at temperature T and has arbitrary spectral density, a non-Markovian
Langevin equation describing the time-evolution of the system particle. 

Relatedly, Chou et al. \cite{Chou2008} have shown that if one replaces
the heavy probe particle of the KZ model with a system of N interacting
identical harmonic oscillators, and if one assumes bilinear coupling
of identical strength between the system and bath position coordinates,
then there exists a canonical transformation that makes it possible
to separate out the CM of the system from its relative degrees of
freedom in the system-bath Hamiltonian. In other words, the transformed
Hamiltonian, in the system degrees of freedom, is of the same form
as the Schr{\"o}dinger Hamiltonian in (36), the latter obtained from
Oriols et al.'s coordinate transformation (12-13). Moreover, the transformed
Hamiltonian entails that only the CM couples to the bath particles.
Under the assumptions that (i) the system and bath are initially uncorrelated,
(ii) the heat bath is initially at thermal equilibrium at temperature
T, and (iii) the spectral density of the bath is arbitrary, Chou et
al. then use the transformed Hamiltonian to define the unitary evolution
of a system-bath density matrix. Tracing over the bath degrees of
freedom, they find that the reduced density matrix for the system
evolves by a non-Markovian master equation of Hu-Paz-Zhang type \cite{HuPazZhang1992}.
Such a master equation is, of course, well-known in the theory of
quantum Brownian motion for open systems \cite{Schlosshauer2008}.

Our proposal, then, is to construct a KZ-type model from ZSM-Newton/Coulomb
(or dBB-Newton/Coulomb), using the same starting assumptions as Chou
et al., and applying the Oriols et al. scheme to the system and bath,
respectively. This should make it possible to show that decoherence
of the system wavefunction via interaction with the bath leads, under
unitary evolution, to a macroscopic superposition of effectively orthogonal
system-bath product states, and that such an evolution is accompanied
by effective collapse of the system-bath configuration into one of
the system-bath product states. In addition, the evolution of the
system's CM particle position, with the bath variables integrated
out, should be described by a non-Markovian modified Langevin equation,
where the modifying terms are the quantum forces from the CM's quantum
kinetic and the quantum kinetics of the relative degrees of freedom,
and where both types of quantum kinetics are constructed from the
effective system wavefunction to which the system configuration has
collapsed. Then, taking the large particle number limits simultaneously
for system and bath, it should be possible to show, by applying the
arguments in section 3 of the present paper, that the equations of
motion for the system and bath CM positions become effectively classical.
In other words, we should recover the classical non-Markovian Langevin
equation for the heavy particle in the classical KZ model.

The details of this proposal will be worked out in a stand-alone paper.

\section{Conclusion}

We have applied Oriols et al.'s large-N-CM approximation scheme to
a system of N identical, non-relativistic, \emph{zbw} particles interacting
via potentials $\hat{U}_{int}(\hat{x}_{j}-\hat{x}_{k})$ and with
external potentials $\hat{U}_{ext}(\hat{x}_{j})$. This made it possible
to: (i) self-consistently describe large numbers of identical \emph{zbw}
particles interacting classical-gravitationally/electrostatically,
without an independent particle approximation; (ii) avoid macroscopic
semiclassical gravitational/electrostatic cat states and recover K-particle
classical Newtonian gravity/electrodynamics for the CM descriptions
of gravitationally/electrostatically interacting macroscopic particles
(where the macroscopic particles are built out of interacting \emph{zbw}
particles); and (iii) recover classical Vlasov-Poisson mean-field
theory for macroscopic particles that interact gravitationally/electrostatically,
in the weak-coupling large K limit. In addition, we have sketched
a proposal for an explicit model of environmental decoherence consistent
with the Oriols et al. large-N-CM approximation scheme, the purpose
of which is to explicitly demonstrate our claim that environmental
decoherence plus effective collapse entails WFPs associated to real-world
macroscopic objects.

We leave for future work the task of extending the ZSM-based large-N-CM
approximation scheme to relativistic massive particles and fields,
in flat and curved spacetimes.

\bibliographystyle{unsrt}
\bibliography{ZSM-Newton}

\end{document}